\RequirePackage{fix-cm}
\documentclass[smallextended]{svjour3} 
\smartqed  
\usepackage{graphicx}
\usepackage{mathtools}
\usepackage{amsmath}
\usepackage{amssymb}
\usepackage[colorlinks]{hyperref}
\usepackage{ae}
\usepackage[T1]{fontenc}
\usepackage[ansinew]{inputenc}
\begin{document}

\title{Local Quantum Uncertainty in Two-Qubit Separable States: A Case Study}
\author{Ajoy Sen         \and
        Debasis Sarkar \and
        Amit Bhar }


\institute{First Author \at
              Department of Applied Mathematics, University of Calcutta, 92, A.P.C. Road, Kolkata-700009, India \\
              \email{ajoy.sn@.com}
            \and
           Second Author \at
           Department of Applied Mathematics, University of Calcutta, 92, A.P.C. Road, Kolkata-700009, India \\
              \email{dsappmath@caluniv.ac.in}
           \and
           Third Author \at
           Department of Mathematics, Jogesh Chandra Chaudhuri College, 30, Prince Anwar Shah Road, Kolkata-700033, India\\
           \email{bhar.amit@yahoo.com}
}

\date{Received: date / Accepted: date}

\maketitle

\begin{abstract}
Recent findings suggest, separable states, which are otherwise of no use in entanglement dependent tasks, can also be used in information processing tasks that depend upon the discord type general  non classical correlations. In this work, we explore the nature of uncertainty in separable states as measured by local quantum uncertainty. Particularly in two-qubit system, we find separable X-state which has maximum local quantum uncertainty. Interestingly, this separable state coincides with the separable state, having maximum geometric discord. We also search for the maximum amount of local quantum uncertainty in separable Bell diagonal states. We indicate an interesting connection to the tightness of entropic uncertainty with the state of maximum uncertainty.
\end{abstract}

\section{Introduction}
As a measure of non classical correlation beyond entanglement, discord\cite{zurek,dakic,modi} has generated lot of interest in recent years. Numerous literature has been engaged in understanding its precise role in both the quantum computing and information theoretic tasks, e.g., DQC1 model \cite{dqc1,dqc2,dqc3}, Grover search algorithm\cite{grover}, remote state preparation\cite{rsp,rsp2}, state merging\cite{merge,merge2} entanglement distribution\cite{entdist,entdist2}, state discrimination\cite{statedis,statedis2,aosd,aosd2}, two-qubit state ordering\cite{ordering}, quantum cryptography\cite{crypto}. Recently, an operational method of using discord, as a resource, has been experimentally established \cite{gu} and an interesting connection between discord and interferometric power of quantum state has been established \cite{interfero}. Several other versions of discord have also been proposed\cite{modi} including geometric and relative entropic discord.

Measurement, in general, disturbs a quantum state. Classically, we can measure any two observable with arbitrary accuracy. However, such kind of measurement is not possible in quantum theory even if we use flawless measurement device. Heisenberg uncertainty principal provides the precession in such kind of measurement. No quantum state shows uncertainty under the measurement of single global observable. However, measurement of a single local observable can manifest uncertainty in a quantum state. Uncertainty in a quantum state can arise due to its classical mixing or due to its non-commutativity with the measuring observable. Girolami \textit{et al.} \cite{girolami} have introduced the concept of local quantum uncertainty (LQU, in short) as a measure of minimum uncertainty by measurement of a single local observable on a quantum state. This quantity identifies the true quantum part of error which arises due to non-commutativity between state and observable and  it does not change under classical mixing. Zero uncertainty implies the existence of quantum certain (commutative) local observable corresponding to the state. Every entangled state possesses this kind of uncertainty, i.e., there is no quantum certain local observable for any entangled state. Even, mixed separable states can show the same characteristic. The only class of states which remain invariant under such local measurement is the states with zero quantum discord\cite{zurek}. Thus the non-zero discord state show uncertainty under the measurement of a single local observable.

For a bi-partite quantum state $\rho_{AB}$, local quantum uncertainty(LQU) is defined as,
\begin{equation}\label{lqu}
\mathcal{U}^\Lambda_A(\rho_{AB}):=\min_{K^\Lambda} I(\rho_{AB}, K^\Lambda)
\end{equation}
The quantity $I$ denotes skew Information, defined by Wigner and Yanase\cite{wigner} as,
\begin{equation}
I(\rho, K):=-\frac{1}{2}\text{tr}\{[\sqrt{\rho},K]^2\}
\end{equation}
and clearly it is a measure of non-commutativity between a quantum state and an observable. The minimization in (\ref{lqu}) is performed over all local \textit{maximally informative observable} (or non-degenerate spectrum $\Lambda$) $K^\Lambda=K_{A}^{\Lambda}\otimes \mathbb{I}$.

Local quantum uncertainty is inherently an asymmetric quantity. It is invariant under local unitary operations. It vanishes for states with zero discord. For pure bi-partite states, it reduces to a entanglement monotone(linear entropy of reduced subsystems). In, two-qubit system all $\Lambda$ dependent quantities become proportional and the dependency on $\Lambda$ can be dropped. So, LQU can be taken as a measure of bi-partite quantumness in two-quit system.

Local quantum uncertainty has been calculated in DQC1 model and it can explain quantum advantages by separable states like discord. LQU also have significant application in quantum metrology. It has close connection to quantum Fisher information\cite{fisher,fisher2} due to the link between Fisher information metric and skew information\cite{fishskew,fishskew2}. It provides a upper bound to the variance of best estimator of the parameter in parameter estimation protocol.
LQU has geometrical significance in terms of Hellinger distance\cite{hellinger} between the state and its least disturbed counterpart after root of unity local unitary operation. More specifically, for a quantum state $\rho$, $\mathcal{U}_A(\rho)=D_{H}(\rho, K^A\rho K^A)$, where $D_{H}^2(\rho,\chi):=\frac{1}{2}\text{tr}\{(\sqrt{\rho}-\sqrt{\chi})^2\}$ is the squared Hellinger distance and $K^A$ is root of unity operation on party A.

Explicit closed form of LQU has been derived only for some symmetric class of states and for simple systems \cite{girolami,ajoy}. For a quantum state $\rho$ of $2\otimes n$ system,
\begin{equation}\label{2lqu}
\mathcal{U}_A(\rho)=1-\lambda_{max}(\mathcal{W})
\end{equation}
where $\lambda_{max}$ is the maximum eigenvalue of the matrix $\mathcal{W}=(w_{ij})_{3\times3}$, $w_{ij}=\text{tr}\{\sqrt{\rho}(\sigma_{i}\otimes \mathbb{I})\sqrt{\rho}(\sigma_{j}\otimes \mathbb{I})\}$ and $\sigma_{i}$'s are standard Pauli matrices in this case. In two-qubit system, Bell states achieve  maximum uncertainty value $1$, which shows that the definition of LQU is normalized. Since, LQU indicates non zero value for separable state, so it will be interesting to investigate on ``\textit{how much we can do with separable state}?" i.e., maximum uncertainty we can achieve with separable state. LQU provides a guaranteed upper bound to the variance of best estimator of parameter in parameter estimation protocol. The maximal value will provide the limit of precession which can be achieved using separable states in such metrological task. Our next curiosity will be whether there is any connection between such separable states with maximum LQU and maximum geometric discord. The problem is in fact a mathematical optimization problem over all separable states $(\mathcal{S})$ as,
\begin{equation}\label{opt}
\begin{split}
\max_{\rho\in \mathcal{S}} \mathcal{U}_A(\rho)&=\max_{\rho\in \mathcal{S}}(1-\lambda_{max}(\mathcal{W}))\\
&=1-\min_{\rho\in \mathcal{S}}(\lambda_{max}(\mathcal{W}))\\
\end{split}
\end{equation}
i.e., we need to find out the minimum of $\lambda_{max}(\mathcal{W})$ over all separable states. We will begin form separable X class of states. This is an important subclass of states as it contains Bell diagonal states and Werner states. This class of states are frequently encountered in studying quantum dynamics, condensed matter systems, etc \cite{xstate1,xstate2,xstate3,xstate4,xstate5}.

\section{LQU in Separable X-state}
\begin{figure}[htb]
  \centering
  \includegraphics[width=3in]{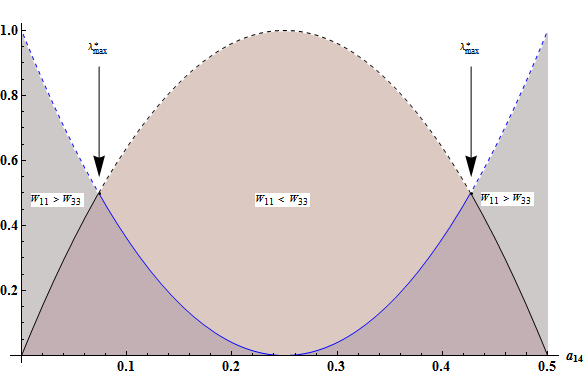}\\
  \caption{(color online) The figure shows the region corresponding to $w_{11}\geq w_{33}$ and $w_{33}\geq w_{11}$. The dotted upper boundary curve indicates the value of  $\lambda_{max}$. The two marked red points indicate the minimum value of $\lambda_{max}$ in those regions and hence the points corresponds to the solution of the optimization problem, i.e., maximum LQU }\label{compare}
\end{figure}

Here we will start with the standard way of parameterizing a two-qubit X state. Let us, first, consider any two-qubit  state $\rho$ in the form,
\begin{equation}\label{xstate}
\rho=\left(
  \begin{array}{cccc}
    a_{11} & 0 & 0 & a_{14} \\
    0 & a_{22} & a_{23} & 0 \\
    0 & a_{32} & a_{33} & 0 \\
    a_{41} & 0 & 0 & a_{44} \\
  \end{array}
\right)
\end{equation}
The elements of the matrix satisfy,
\begin{eqnarray}
\begin{rcases}
\begin{split}
&a_{14}=a_{41}^{\dagger},\quad a_{23}=a_{32}^{\dagger} \quad\text{(Complex Conjugation)}\\
&\sum_{i=1}^4 a_{ii}=1,\quad \text{(Normalization)}\\
&a_{11}a_{44}\geq a_{14}^2,\quad a_{22}a_{33}\geq a_{23}^2,\quad \text{(Positivity)}\\
\end{split}
\end{rcases}
\end{eqnarray}
The state contains seven independent parameter. However, LQU is local unitary invariant and we can easily drive out the phases from off diagonal element by mere local unitary operation. Hence, we are left with only five positive real parameters and henceforth with out loss of generality we will consider all $a_{ij}$'s as real and non negative. This X state has four real eigenvalues $\lambda_{0}$, $\lambda_{1}$, $\lambda_{2}$, $\lambda_{3}$ and corresponding eigenvectors are $|v_{0}^{'}\rangle $, $|v_{1}^{'}\rangle $, $|v_{2}^{'} \rangle $, $|v_{3}^{'} \rangle $. Here, $\lambda_{i}$'s and $|v_{i}^{'}\rangle $'s have dependence on $a_{ij}$'s. Thus, $\rho$ can be decomposed as $\rho=\sum_{i=0}^{3} \lambda_{i}|v_{i}\rangle\langle v_{i}|$ with $|v_{i}\rangle$ being the normalized form of $|v_{i}^{'}\rangle$. The state vectors $|v_{i}\rangle$'s  are mutually orthonormal and we can easily write $\sqrt{\rho}=\sum_{i=0}^{3} \sqrt{\lambda_{i}}|v_{i}\rangle\langle v_{i}|$. After a bit simplification it reads,
\begin{equation}\label{root}
\begin{split}
\sqrt{\rho}=\left(
  \begin{array}{cccc}
    \alpha_{1} & 0 & 0 & \alpha_{5} \\
    0 & \alpha_{2} & \alpha_{6} & 0 \\
    0 & a_{6} & \alpha_{3} & 0 \\
    \alpha_{5} & 0 & 0 & \alpha_{4} \\
  \end{array}
\right)\\
\end{split}
\end{equation}
with
\begin{eqnarray}
\begin{rcases}
\begin{split}
\alpha_{1}&=\left(\frac{\sqrt{\lambda_{0}}\omega_{0}^2}{\omega_{0}^2+1}+\frac{\sqrt{\lambda_{1}}\omega_{1}^2}{\omega_{1}^2+1}\right)\\
\alpha_{2}&=\left(\frac{\sqrt{\lambda_{2}}\omega_{2}^2}{\omega_{2}^2+1}+\frac{\sqrt{\lambda_{3}}\omega_{3}^2}{\omega_{3}^2+1}\right)\\
\alpha_{3}&=\left(\frac{\sqrt{\lambda_{2}}}{\omega_{2}^2+1}+\frac{\sqrt{\lambda_{3}}}{\omega_{3}^2+1}\right)\\
\alpha_{4}&=\left(\frac{\sqrt{\lambda_{0}}}{\omega_{0}^2+1}+\frac{\sqrt{\lambda_{1}}}{\omega_{1}^2+1}\right)\\
\alpha_{5}&=\left(\frac{\sqrt{\lambda_{0}}\omega_{0}}{\omega_{0}^2+1}+\frac{\sqrt{\lambda_{1}}\omega_{1}}{\omega_{1}^2+1}\right)\\
\alpha_{6}&=\left(\frac{\sqrt{\lambda_{2}}\omega_{2}}{\omega_{2}^2+1}+\frac{\sqrt{\lambda_{3}}\omega_{3}}{\omega_{3}^2+1}\right)\\
\end{split}
\end{rcases}
\end{eqnarray}
and
\begin{eqnarray}
\begin{rcases}
\begin{split}
\omega_{0}&=\left(\frac{a_{11}-a_{14}+\lambda_{0}-\lambda_{1}}{2a_{14}}\right)\\
\omega_{1}&=\left(\frac{a_{11}-a_{14}-\lambda_{0}+\lambda_{1}}{2a_{14}}\right)\\
\omega_{2}&=\left(\frac{a_{22}-a_{33}+\lambda_{2}-\lambda_{3}}{2a_{23}}\right)\\
\omega_{3}&=\left(\frac{a_{11}-a_{14}-\lambda_{2}+\lambda_{3}}{2a_{23}}\right)\\
\end{split}
\end{rcases}
\end{eqnarray}
We also have the relation $\langle v_{i}^{'}|v_{i}^{'}\rangle=\omega_{i}^{2}+1$. According to the definition of LQU, and it turns out that, $\mathcal{W}=\text{Diag}(2(\alpha_{1}\alpha_{3}+\alpha_{2}\alpha_{4}+\alpha_{5}\alpha_{6}),
2(\alpha_{1}\alpha_{3}+\alpha_{2}\alpha_{4}-\alpha_{5}\alpha_{6}),\sum_{i=1}^{6}\alpha_{i}^{2}
-3\alpha_{5}^{2}-3\alpha_{6}^{2})$. Let us define,
\begin{eqnarray}
\begin{rcases}
\begin{split}
w_{11}&=2\left(\alpha_{1}\alpha_{3}+\alpha_{2}\alpha_{4}+\alpha_{5}\alpha_{6}\right)\\
w_{22}&=2\left(\alpha_{1}\alpha_{3}+\alpha_{2}\alpha_{4}-\alpha_{5}\alpha_{6}\right)\\
w_{33}&=\left(\sum_{i=1}^{6}\alpha_{i}^{2}-3\alpha_{5}^{2}-3\alpha_{6}^{2}\right)\\
\end{split}
\end{rcases}
\end{eqnarray}
Till now, we have not taken into account the separability condition. PPT (Positive partial transpose) criteria works as a necessary and sufficient condition of separability in two-qubit system. PPT criteria gives the following two separability conditions,
\begin{equation}\label{1}
a_{11}a_{44}\geq a_{23}^2 \qquad \text{and}\qquad  a_{22}a_{33}\geq a_{14}^2
\end{equation}
These two conditions are dual to the original positivity constraints. We can write an optimization problem of LQU over all separable X-states as,
\begin{equation}\label{PP1}
\begin{split}
\text{Minimize}\quad & \lambda_{max}(\mathcal{W})=\max\{w_{11},w_{22},w_{33}\}\\
\text{Subject to}:\\
&a_{14}\leq \min(\sqrt{a_{11}a_{44}},\sqrt{a_{22}a_{33}})\\
&a_{23}\leq \min(\sqrt{a_{11}a_{44}},\sqrt{a_{22}a_{33}})\\
&\sum a_{ii}=1\\
&a_{ij}\geq 0, \quad \forall i,j
\end{split}
\end{equation}
If $\lambda_{max}^{*}$ is the solution of the minimization problem, we will have maximum LQU,  $\mathcal{U}_A^{*}=\text{max}(1-\lambda_{max})=1-\lambda_{max}^{*}$.
Now let us consider the case $\alpha_{5}\alpha_{6}\geq 0$. In this case, $w_{11}\geq w_{22}$. Hence $\lambda_{max}(\mathcal{W})=\max\{w_{11},w_{33}\}$. Depending upon the sign of $w_{11}-w_{33}$ we formulate two optimization problems from (\ref{PP1}) as,
\begin{equation}\label{PP2}
\begin{split}
\text{Minimize}\quad & \lambda_{max}(\mathcal{W})=w_{33}\\
\text{Subject to}:\\
&w_{11}\leq w_{33}\\
&\alpha_{5}\alpha_{6}\geq 0\\
&a_{14}\leq \min(\sqrt{a_{11}a_{44}},\sqrt{a_{22}a_{33}})\\
&a_{23}\leq \min(\sqrt{a_{11}a_{44}},\sqrt{a_{22}a_{33}})\\
&\sum a_{ii}=1\\
&a_{ij}\geq 0, \quad \forall i,j
\end{split}
\end{equation}
and
\begin{equation}\label{PP3}
\begin{split}
\text{Minimize}\quad & \lambda_{max}(\mathcal{W})=w_{11}\\
\text{Subject to}:\\
&w_{33}\leq w_{11}\\
&\alpha_{5}\alpha_{6}\geq 0\\
&a_{14}\leq \min(\sqrt{a_{11}a_{44}},\sqrt{a_{22}a_{33}})\\
&a_{23}\leq \min(\sqrt{a_{11}a_{44}},\sqrt{a_{22}a_{33}})\\
&\sum a_{ii}=1\\
&a_{ij}\geq 0, \quad \forall i,j
\end{split}
\end{equation}
In the first case (\ref{PP2}), after simplifying, we get $\mathcal{U}_A=4 \max\{\alpha_{5}^{2}+\alpha_{6}^{2}\}$ and in the second case (\ref{PP3}),  $\mathcal{U}_A=\max\{(\alpha_{1}-\alpha_{3})^2+(\alpha_{2}-\alpha_{4})^2+2(\alpha_{5}-\alpha_{6})^2\}$. A little simple calculation and the form of constraints in the optimization problem suggest us to choose $a_{11}=a_{22}$, $a_{33}=a_{44}$ and $a_{14}=a_{23}=\sqrt{a_{11}a_{33}}$ for the sake of minimization purpose and we get $w_{11}=4(a_{11}-a_{33})^2$ and $w_{33}=16 a_{11}a_{33}$. We also need to consider normalization condition $a_{11}+a_{33}=\frac{1}{2}$. Under these constraints, after solving, we get $a_{11}=\frac{\sqrt{2}+1}{4\sqrt{2}}$, $a_{33}=\frac{\sqrt{2}-1}{4\sqrt{2}}$. The regions corresponding to the two optimization problem are shown in (\ref{compare}).
Exactly similar analysis follows if we choose $\alpha_{5}\alpha_{6}<0$. In this case we get $a_{11}=\frac{\sqrt{2}-1}{4\sqrt{2}}$, $a_{33}=\frac{\sqrt{2}+1}{4\sqrt{2}}$. The sates corresponding to both the solutions are merely connected by local unitary $\sigma_{x}\otimes \sigma_{x}$. Hence we obtain a unique (up to local unitaries) rank-2  separable X-state $\rho^{*}$ with $\mathcal{U}_A^{*}=\frac{1}{2}$.
\begin{equation}\label{state}
\rho^{*}=\frac{1}{4}\left(
  \begin{array}{cccc}
    \frac{\sqrt{2}+1}{\sqrt{2}} & 0 & 0 & \frac{1}{\sqrt{2}} \\
    0 & \frac{\sqrt{2}+1}{\sqrt{2}} & \frac{1}{\sqrt{2}} & 0 \\
    0 & \frac{1}{\sqrt{2}} & \frac{\sqrt{2}-1}{\sqrt{2}} & 0 \\
    \frac{1}{\sqrt{2}} & 0 & 0 & \frac{\sqrt{2}-1}{\sqrt{2}} \\
  \end{array}
\right)
\end{equation}
Interestingly, exactly  same state (up to local unitaries) was shown\cite{rana} to have maximum geometric discord among separable X-states.

Any two-qubit quantum state, under local unitary equivalence, can be taken as,
$\rho=\frac{1}{4}(\mathbb{I}_2\otimes \mathbb{I}_2 + \mathbf{x^t}\mathbf{\sigma}\otimes \mathbb{I}_2+\mathbb{I}_2\otimes \mathbf{y^t}\mathbf{\sigma}+\sum_{i=1}^{3} t_{i}\sigma_{i}\otimes \sigma_{i})$
where $\mathbb{I}_n$ denotes the identity matrix of order $n$, $\sigma=(\sigma_1,\sigma_2,\sigma_3)$ where $\sigma_i$'s  are usual Pauli matrices, $T=\text{Diag}[t_{1},t_{2},t_{3}]$ is the correlation matrix with components $t_i=\text{tr}(\rho\sigma_i\otimes\sigma_i)$. $\mathbf{x}=(x_1,x_2,x_3)$, $\mathbf{y}=(y_1,y_2,y_3)$ are Bloch vectors with $x_i=\text{tr}(\rho\sigma_i\otimes \mathbb{I}_2)$, $y_i=\text{tr}(\rho\mathbb{I}_2\otimes\sigma_i)$.
Our numerical simulation with separable states did not reveal any state with LQU greater than $\frac{1}{2}$. This tempted us to conjecture that \textit{maximum value of LQU for two-qubit separable states is} $\frac{1}{2}$. This is in same spirit to the similar conjecture on discord, made in \cite{gharibian}. Starting from a different measure (discriminating strength) Farace \textit{et al.}\cite{farace} find a relation between their correlation measure and LQU. For two-qubit system, they considered rank 2, 3 ,4 separable states and performed similar optimization. Their numerical result shows that rank-$2$ states achieve the maximum LQU for B-92 states but the analytical proof is still absent. However they presented analytical proof in other dimensions.\\

\textit{Maximum LQU for separable Bell Diagonal states}: This class of states belongs to X class of states and have the form,
$$\rho=p_{\tiny{I}}|\phi^{+}\rangle\langle \phi^{+}|+p_{x}|\phi^{-}\rangle\langle \phi^{-}|+p_{y}|\psi^{+}\rangle\langle \psi^{+}|+p_{z}|\psi^{-}\rangle\langle \phi^{-}|$$
These states are entangled if any one parameter among $p_{I}$, $p_{x}$, $p_{y}$, $p_{z}$ is greater than $\frac{1}{2}$. In terms of Bloch representation, the state can also be written as,
$$\rho=\frac{1}{4}[\mathbb{I}_2\otimes \mathbb{I}_2+\sum_{i=1}^{3} t_{ii}\sigma_i\otimes\sigma_i]$$
Bloch vectors corresponding to this class can be obtained as $\mathbf{x}=\mathbf{0}$, $\mathbf{y}=\mathbf{0}$.  Whenever $t_{11}=t_{22}=t_{33}=t$ (say) i.e., $T=t \mathbb{I}_3$, LQU corresponding to this class is,
$$\mathcal{U}_A(\rho)=1-\frac{1}{2}\sqrt{1+t}\left(\sqrt{1+t}+\sqrt{1-3t}\right)$$
Maximum value of LQU can reach  $\frac{1}{3}$ in separable domain ($-\frac{1}{3}\leq t\leq \frac{1}{3}$). If we consider any two of the $t_{ii}$ are equal or when all $t_{ii}$'s are unequal, our numerical suggest that we can't reach more than $\frac{1}{3}$ by separable Bell diagonal states.\\

\textit{Dissonance}: $\rho^* \equiv\rho^*_{AB}$ is a rank-$2$ separable state and it can be written as, $\rho^*_{AB}=\lambda_0|\phi_0\rangle\langle\phi_0|+\lambda_0|\phi_0\rangle\langle\phi_0|$ where $|\phi_0\rangle=a_0|00\rangle+b_0|11\rangle$, $|\phi_1\rangle=a_1|10\rangle+b_1|01\rangle$ are orthogonal states and the parameters are $\lambda_0=\lambda_1=\frac{1}{2}$, $a_0=b_1=\frac{\sqrt{2}+1}{\sqrt{4+2\sqrt{2}}}$, $a_1=b_0=\frac{1}{\sqrt{4+2\sqrt{2}}}$. Its dissonance ($D_A$) can be easily obtained\cite{shi} form the purified Koashi-Winter relation,
\begin{equation}
D_A(\rho^*_{AB})=S(\rho^*_A)-S(\rho^*_{AB})+E_F(\rho^*_{BC})
\end{equation}
$E_F$ denotes the entanglement of formation, $S$ denotes von Neumann  entropy and dissonance is considered w.r.t. the measurement on party A.  Let $|\Psi^*_{ABC}\rangle=\sqrt{\lambda_0}|\phi_0\rangle|0\rangle+\sqrt{\lambda_1}|\phi_1\rangle|0\rangle$  be a purification of the state $\rho^*_{AB}$. We can easily evaluate the reduced states $\rho^*_{AB}$, $\rho^*_{BC}$, $\rho^*_{A}$ and  obtain $E_F(\rho^*_{BC})=S(\rho^*_A)\approx0.6$, $S(\rho^*_{AB})=1$. Hence, $D_A(\rho^*_{AB})=0.20175$. This value is in fact very large within separable states and it is higher than the dissonance of Werner  class of states.

\section{Behavior of maximal uncertain separable X-state in Entropic Uncertainty Relation}
The role of maximally uncertain separable state can be investigated in connection to the tightness of entropic uncertainty relation.
In the presence of quantum memory, entropic uncertainty relation was proposed by Berta \textit{et al.} \cite{berta} as,
\begin{equation}
S(P|B)+S(Q|B)\geq -2 \log_2 c(P,Q)+S(A|B)
\end{equation}
and later it was improved by Pati \textit{et al.} \cite{pati} as,
\begin{equation}
\begin{split}
S(P|B)+S(Q|B)\geq -2 \log_2 c(P,Q)+S(A|B)+\\\max{\{0, D_A(\rho_{AB})-J_A(\rho_{AB})\}}
\end{split}
\end{equation}
where $\rho_{AB}$ is the initial state between quantum system $A$ and quantum memory $B$. $S(P|B)$ and $S(Q|B)$ are the conditional von Neumann entropies of the state $\rho_{AB}$ after measurement of the observable $P$ and $Q$ on A respectively and $S(A|B):=S(\rho_{AB})-S(\rho_B)$ is the quantum conditional entropy without measurement. $c(P,Q)\equiv \max_{i,j}{|\langle p_i|q_j\rangle|}$ and $\{p_i\}$, $\{q_j\}$ are the eigenvectors of the observables $P$ and $Q$. $D_A(\rho_{AB})$, $J_A(\rho_{AB})$ denotes quantum discord and classical correlation of the state $\rho_{AB}$ respectively and they are defined as,
\begin{eqnarray}
J_A(\rho_{AB})&=&\max_{\{\Pi_j^A\}} \left[S(\rho_B)-S(\rho_{B|A})\right]\\
D_A(\rho_{AB})&=&I(\rho_{AB})-J_A(\rho_{AB})
\end{eqnarray}
maximum is taken over all projective measurement $\{\Pi_j^A\}$ on party A, $S(\rho_{B|A})$ denotes the standard conditional entropy of the state obtained by the projective measurement on A\cite{zurek}.

\begin{figure}[t]
  \centering
  \includegraphics[width=3in]{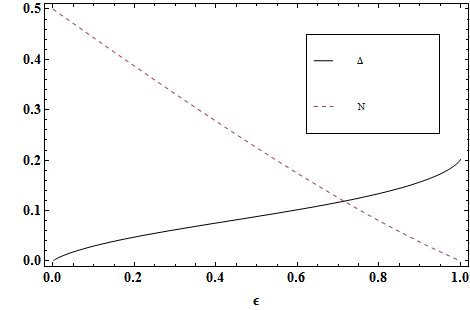}
  \caption{Nature of entanglement as measured by negativity $N$ and uncertainty gap $\Delta$ for the states $\chi$. Both curve shows monotonic behavior. In fact maximum uncertainty gap is achieved at $\epsilon=1$. Uncertainty gap increases as $\epsilon\rightarrow 1$ but negativity decreases. At $\epsilon \approx 0.714$, both becomes equal and then negativity further decays to zero but $\Delta$ increases up to its maximum. }\label{fig2}
\end{figure}

We define the sum of uncertainty $S(P|B)+S(Q|B)$ as $U_B^{P,Q}$ and the lower bound $-2 \log_2 c(P,Q)+S(A|B)+\max{\{0, D_A(\rho_{AB})-J_A(\rho_{AB})\}}$ as $L_B^{P,Q}$. The difference $\Delta^{P,Q}:= U_B^{P,Q}-L_B^{P,Q}$ denotes a kind of uncertainty gap in a quantum state corresponding to the pair of observables $P$ and $Q$. This quantity is obviously non-negative and may characterizes the discrepancy between uncertainty of the measurement outcomes of $P$ and $Q$ \cite{ming}.  Maximally entangled states  have $\Delta^{\sigma_x,\sigma_z}=0$ corresponding to the maximally unbiased spin observables $\sigma_x$ and $\sigma_z$. We are interested in checking the status of this discrepancy around the state of maximal uncertainty. We will consider both separable and entangled states around $\rho^*$. We consider the mixed entangled state $\chi=\epsilon \rho^{*}+(1-\epsilon)|\phi^{+}\rangle \langle\phi^{+}|$, where $|\phi^{+}\rangle=\frac{|00\rangle+|11\rangle}{\sqrt{2}}$. For this state, we have the conditional entropies $S(Q|B)=\mathcal{H}(\{\frac{(2-\sqrt{2})\epsilon}{8},\frac{(2+\sqrt{2})\epsilon}{8},\frac{4-(2+\sqrt{2})\epsilon}{8},\frac{4-(2-\sqrt{2})\epsilon}{8}\})-1$ and $S(P|B)=\mathcal{H}(\{\frac{(2-\sqrt{2})\epsilon}{8},\frac{(2-\sqrt{2})\epsilon}{8},\frac{4-(2-\sqrt{2})\epsilon}{8},\frac{4-(2-\sqrt{2})\epsilon}{8}\})-1$ corresponding to spin observables $\sigma_x$ and $\sigma_z$. $\mathcal{H}(\{p_i\})$ is the usual Shannon entropy of the probability distribution $\{p_i\}$. The uncertainty gap is monotonic as evident from the FIG. \ref{fig2}. $\Delta^{\sigma_x,\sigma_z}$ obtains its highest value corresponding to $\epsilon=1$, i.e., $\rho^{*}$ has maximum uncertainty gap from this class. Even if we consider the noisy mixed separable state $p\rho^{*}+\frac{1-p}{4}\mathbb{I}$, the state $\rho^{*}$ shows the highest discrepancy.

\section{Conclusion}
We have thus shown that among separable X states there is a unique state which attains the maximum value of LQU. The same state also attains the maximum value of geometric discord among similar class of states. We believe that this result can be extended (based on our numerical exploration and also from the work of Farace et. al.) to whole separable class of states and in that case this value will be a good indicator of entanglement since the amount of uncertainty beyond the value $\frac{1}{2}$ necessarily imply the existence of entanglement. We hope, our result will provide a limit that we can achieve by separable states in some other quantum information theoretic tasks or some protocols in near future. 

\begin{acknowledgements}
The author A. Sen acknowledges the financial support from University Grants Commission, New Delhi, India. The author D. Sarkar also acknowledges DST SERB for financial support.
\end{acknowledgements}

\end{document}